\documentclass[twocolumn, twocolappendix]{aastex631}
\usepackage{amsmath}
\usepackage{xcolor}

\usepackage{apjfonts}
\usepackage{float}
\usepackage{enumitem}

\begin{document}

\title{Coincident Multimessenger Bursts from Eccentric Supermassive Binary Black Holes}

\correspondingauthor{Vikram Manikantan}
\email{vik@arizona.edu}

\author[0000-0003-0547-6158]{Vikram Manikantan}
\affiliation{Steward Observatory \& Department of Astronomy, University of Arizona, Tucson, AZ 85721, USA}

\author[0000-0002-8099-9023]{Vasileios Paschalidis}
\affiliation{Steward Observatory \& Department of Astronomy, University of Arizona, Tucson, AZ 85721, USA}
\affiliation{Department of Physics, University of Arizona, Tucson, AZ 85721, USA}

\author[0000-0003-3696-6408]{Gabriele Bozzola}
\affiliation{Steward Observatory \& Department of Astronomy, University of Arizona, Tucson, AZ 85721, USA}
\affiliation{Division of Geological and Planetary Sciences, California Institute of Technology, Pasadena, 91125 California, USA}

\begin{abstract}
Supermassive binary black holes are a key target for the future Laser Interferometer Space Antenna and excellent multi-messenger sources across the electromagnetic spectrum. However, unique features of their electromagnetic emission (EM) that are needed to distinguish them from single supermassive black holes are still being established. Here, we conduct the first magnetohydrodynamic simulation of disk accretion onto equal-mass, nonspinning, eccentric binary black holes in full general relativity, incorporating synchrotron radiation transport through the dual jet in postprocessing. Focusing on a binary in the strong-field dynamical spacetime regime with eccentricity $e=0.3$ as a point of principle, we show that the total accretion rate exhibits periodicity on the binary orbital period. We also show, for the first time, that this periodicity is reflected in the jet Poynting luminosity {\it and} the optically thin synchrotron emission from the jet base. Furthermore, we find a distinct EM signature for eccentric binaries: they spend more time in a {\it low} emission state (at apocenter) and less in a {\it high} state (at pericenter). Additionally, we find that the eccentric binary quasiperiodic gravitational-wave (GW) bursts are coincident with the bursts in Poynting luminosity and synchrotron emission. Finally, we discuss how multimessenger EM and GW observations of these systems can help probe plasma physics in their jet.

\end{abstract}

\keywords{Black Holes (162) -- Gravitational Waves (678) -- High-energy Astrophysics (739) -- Magnetohydrodynamical Simulations (1966)}

\section{Introduction} \label{sec:intro}

The inspiral and merger of supermassive binary black holes (SMBBHs) are main targets for the future space-based gravitational wave (GW) Laser Interferometer Space Antenna (LISA)~\citep{LISA_white_paper, lisa_gw_whitepaper, amaro-seoane_astrophysics_2023}. At least a fraction of SMBBHs are expected to exist in hot gas~\citep{barnes_formation_2002, rodriguez_2009, chen_enhanced_2009, li_direct_2019}. This makes them ideal for multimessenger astronomy because, in addition to GWs, gas accretion will drive emission across the electromagnetic (EM) spectrum~\citep[see][for a recent review]{bogdanovic_bhb_review}. Pulsar Timing Arrays (PTAs) also target SMBBHs, but unless their sensitivity increases significantly~\citep{rosado_expected_2015,nanograv, kelley_single_2018,aggarwal_nanograv_2019,babak_forecasting_2024}, detecting GWs from individual SMBBHs may have to wait until the mid 2030s for the launch of LISA.

Identifying EM signals that are unique to SMBBHs is essential to maximizing the scientific yield of multimessenger astronomy~\citep{Schnittman_2011,bogdanovic_bhb_review}, as coincident observations of EM and GW signals can probe fundamental physics, gravity, astrophysics and cosmology~\citep{LISA_white_paper, lisa_gw_whitepaper, lisa_multimessenger_whitepaper, arun_new_2022, amaro-seoane_astrophysics_2023}. Over 200 SMBBH candidates have been identified through sky surveys~\citep{rodriguez_2006, rodriguez_2009, charisi_multiple_2015, graham_systematic_2015, panstars_smbh, oneill_unanticipated_2022, kiehlmann_pks_2024}, including more than 25 inferred to be in the strong-field dynamical regime~\citep[see Fig. 1 in][]{bright_minidisk}. To evaluate these potentially relativistic candidates, we must use theoretical models of SMBBH accretion to predict their smoking-gun EM signatures.
    
To reliably model SMBBHs from first principles during their late inspiral and merger requires simulations in full general relativity (GR) coupled to magnetohydrodynamics (MHD), microphysics, and radiation. Performing such simulations with existing computational resources and numerical methods is not currently feasible due to the vast range of length and timescales involved. So, to make progress at this time, we have to make simplifying assumptions. Some methods forego dynamical relativistic gravity~\citep{Most:2024qus,Most:2024onq}, and use 2D simulations, \citep[see, e.g.,][and references therein for recent work]{munoz_circumbinary_2020, westernacher-schneider_multi-band_2022, lai_circumbinary_2023, delaurentiis_relativistic_2024}. Others adopt post-Newtonian background metrics \citep[see, e.g.,][and references therein for recent work]{avara_accretion_2023, porter_parameter_2024}. Finally, fully general relativistic $3+1$ approaches solve the Einstein equations without approximation~\citep[see][for recent reviews]{gold_relativistic_2019,Cattorini:2023akr}. To date, no studies have treated microphysics or full radiation transport in $3+1$ GRMHD simulations.

Recent studies of accretion onto BBHs in full $3+1$ GR have investigated various parameters~\citep{khan_disks_2018, gold_accretion_2014, paschalidis_minidisk_2021, bright_minidisk, cattorini_misaligned_2022, ruiz_unequal_2023, fedrigo_grmhd_2024}. However, accretion onto eccentric SMBBHs in full GR remains unexplored. Treating eccentricity is important, as an increasing number of works show that disk-SMBBH interactions can excite substantial binary orbital eccentricity~\citep{roedig_limiting_2011,siwek_2024,valli_long-term_2024,franchini_behaviour_2024}. 

Although GWs tend to circularize orbits as the binary inspirals~\citep{peters_matthews}, non-negligible eccentricity can remain down to the late inspiral depending on disk and binary parameters. Assuming thin disks, $H/R\sim \mathcal{O}(0.01)$, BBHs with mass $M < 10^6 M_\odot$, and mass ratio $q\lesssim0.1$, the eccentricity in the LISA band can potentially be ${\cal O}(0.1)$~\citep{roedig_limiting_2011}. However, the residual eccentricity in the LISA band depends (among other things) on the binary-disk decoupling radius, which shrinks with increasing $H/R$~\citep{farris_binary_2012}. For thin disks, this radius can be $\mathcal{O}(200M)$, in which case the BBH can radiate away most of its eccentricity before or as it enters the LISA band. For thicker disks ($H/R\sim 0.1$, relevant, e.g., for slim disks~\citep{SlimDisksAbramovich}), the decoupling radius can be as small as $30M$~\citep{farris_binary_2012}, and the matter torque on the binary can be maintained for longer, allowing the binary to maintain higher eccentricity as it enters the LISA band. Therefore, it remains an open question for what range of BBH mass and mass ratio binaries in a slim disk environment can achieve high-eccentricity as they enter the LISA band. Additionally, chaotic non-hierarchical three-body interactions - which can arise following a triple galaxy merger~\citep{2021arXiv210612441Y,Ni_2022} - can excite orbital eccentricity up to order unity even at relativistic separations~\citep{ryu_multiple_bh}.

Moreover, eccentric binaries probe more relativistic velocities than quasi-circular ones (at the same orbital period), and even one high-eccentricity binary detection could provide a wealth of opportunities for probing extreme gravity and astrophysics. Therefore, studying theoretically high eccentricity binaries in the dynamical spacetime regime is very important.

In this work, we present the first $3+1$ full GR magnetohydrodynamic simulation of circumbinary disk accretion onto a BBH with initial eccentricity $e = 0.3$, and perform the first synchrotron radiation transport calculation through the dual jet in postprocessing. While the value of eccentricity we use here may be high by traditional astrophysical expectations for comparable-mass binaries, it provides the basis for our point-of-principle calculations and addresses three key questions: i) how does orbital eccentricity impact the mass accretion rate and its periodicity? ii) how does it affect the Poynting luminosity and synchrotron emission? iii) what is its impact on the multimessenger picture of these sources?
    
This paper is structured as follows: in Section \ref{sec:approach} we describe the methods we adopt; in Section~\ref{sec:results} we describe the accretion flow onto the binary and jet launching; in Section~\ref{sec:synchrotron} we report the results of our radiative transfer calculation of synchrotron emission within the jet and discuss its potential detectability and the simultaneous GW and EM emission from our models, and in Section \ref{sec:conc} we conclude with a discussion of our findings. Unless otherwise stated, we adopt geometrized units in which $G = c = 1$.

\section{Methods} \label{sec:approach}

    \subsection{Initial data}
        \paragraph{Spacetime} We use the {\tt TwoPunctures} thorn to generate the spacetime initial data for an equal-mass, non-spinning, eccentric BBH~\citep{ansorg_single-domain_2004,Paschalidis:2013oya}. The BHs are initialized at apocenter with coordinate separation $d/M \sim 26$. We introduce orbital eccentricity by first computing the 3rd-order Post-Newtonian linear momenta corresponding to a quasi-circular BBH, and then adjusting their tangential component by a factor of $\sqrt{1-e}$, where $e$ is our target eccentricity. In a follow-up paper we will provide the details on how we measure the binary eccentricity from our simulations, along with comparisons with additional eccentricities and with a quasi-circular binary~\citep{Manikantan_BBH_eccentric}.
    
        \paragraph{Matter} We adopt the power-law torus solution for the matter initial conditions as previously described in \citep{gold_accretion_2014,khan_disks_2018}. We set the inner edge of the circumbinary disk (CBD)  at $r/M = 18$ with specific angular momentum of $l = 5.15$ and disk outer edge at $r/M \simeq 100$. We use a $\Gamma$-law equation of state, with  $\Gamma = 4/3$ -- appropriate for radiation dominated disks. We seed the disk with a poloidal magnetic field as in~\citep{khan_disks_2018}. The initial magnetic field renders the CBD unstable to the magnetorotational instability (MRI) \citep{Balbus}. 
    
    \subsection{Evolution}
    
        \paragraph{Spacetime} We evolve the spacetime by solving full Einstein equations in the Baumgarte-Shapiro-Shibata-Nakamura (BSSN) formalism of general relativity \citep{Shibata1995, Baumgarte1998} as implemented in the \verb|LeanBSSN| code using 6th-order finite differences~\citep{sperhake_binary_2007}. We adopt the moving puncture gauge conditions \citep{baker_gravitational-wave_2006, campanelli_accurate_2006} with the shift vector parameter $\eta$ set to $\eta=1.4/M$. 
    
        \paragraph{Matter} We employ the 3D general relativistic  magnetohydrodynamic (GRMHD), adaptive-mesh-refinement (AMR) \verb|IllinoisGRMHD| code \citep{ilgrmhd} within the \verb|Einstein Toolkit| \citep{einsteintoolkit}, which employs the Cactus/Carpet infrastructure \citep{Goodale2002a, schnetter_carpet_2016}. \verb|IllinoisGRMHD| evolves the equations of ideal magnetohydrodynamics in flux-conservative form via the HLL Riemann solver \citep{toro_riemann_2009}, and the Piecewise-Parabolic reconstruction \citep{colella_piecewise_1984}. These methods have been described in \cite{ilgrmhd} and have been extensively tested against other codes in \citep{porth_event_2019}. For our EM gauge choice for the vector potential formulation of the induction equation, we use the generalized Lorenz gauge condition of~\citep{Etienne:2011re,farris_binary_2012} and set the Lorenz gauge damping parameter to $\xi=8/M$. Finally, the fluid does not back react onto the spacetime, since the spacetime mass/energy content is dominated by the SMBBH.
        
        \paragraph{Grid} We adopt \verb|Carpet| \citep{schnetter_carpet_2016} for adaptive mesh refinement (AMR). We use a three-dimensional Cartesian grid with the outer boundary extending from $-5120 \, M$ to $+5120 \, M$ in the $x, y,$ and $z$ directions, with a total of 14 refinement levels. We have 3 sets of nested AMR boxes, one centered on the center-of-mass, and two centered on each of the two BHs in the binary. The half side length of an AMR level $i$ is $5120\times 2^{-(i-1)}M$, $i=1,\ldots 14$. The grid spacing on the coarsest (finest) refinement level is  $\Delta x = 128 \, M$ ($\Delta x=M/64$). We do not treat radiative feedback, heating, or cooling. We resolve the fastest growing mode MRI wavelength with at least 20 zones in the disk and a maximum of 50 zones at the inner edge of the disk.
    
        \paragraph{Diagnostics} We adopt the same diagnostic tools as in \cite{bright_minidisk} to measure the rest-mass accretion rate ($\dot{M}$) and outgoing Poynting flux. The latter we compute on a sphere of coordinate radius $200 \, M$ from the binary center of mass, thereby encompassing the entire BBH-disk system.
        
        We locate apparent horizons with \texttt{AHFinderDirect} \citep{Thornburg2004}. We perform all Fourier analysis with the \verb|scipy.fft| function \citep{scipy}. We measure the orbital frequency of the binary with the Fourier transform of the first time derivative of the unfolded phase ($\varphi$) of the $\ell, m=(2,2)$ mode, $f_{22}\equiv\frac{d\varphi}{dt}$, which we extract using the \texttt{NPScalars} thorn part of Canuda suite \citep{canuda,canudacode}. We use the package \verb|kuibit| \citep{kuibit} for all our analysis.

    \subsection{Synchrotron modeling}
    
        To model synchrotron emission, we solve the radiation transfer equation (1.23) of~\cite{rybicki_lightman}, using the synchrotron emissivity given by Eq. (6.36) of~\cite{rybicki_lightman} multiplied by $1/4\pi$, i.e., we average out the distribution over solid angle. Moreover, in the emissivity we average the pitch angle, $\alpha$, out of the equation by assuming that the electrons follow an isotropic pitch angle distribution between $0<\alpha<\pi/2$. We adopt the synchrotron self-absorption coefficient given by Eq. (6.53) of~\cite{rybicki_lightman}. This emissivity and absorption coefficient corresponds to a power-law electron distribution $N(E) \, dE = C E^{-p} \, dE$. This distribution has two constants that define it for a given $p$: $C$ and the miminum electron energy $E_{\rm min}$. To compute $C$ we impose charge neutrality with the ion density in our simulations, and we compute $E_{\rm min}$ by adopting a fixed ratio between the electrons and the magnetic field energy density of 10\%. We also set $E_{\rm min} = 2m_e c^2$, where $m_e$ is the electron mass, as an alternative. For thermal synchrotron we follow~\cite{rybicki_lightman, tsouros_energy_2017}.

\section{Results} \label{sec:results}
    \subsection{Accretion Flow} \label{sec:accretion}

    \begin{figure*}[ht]
        \centering
        \includegraphics[width=1\textwidth]{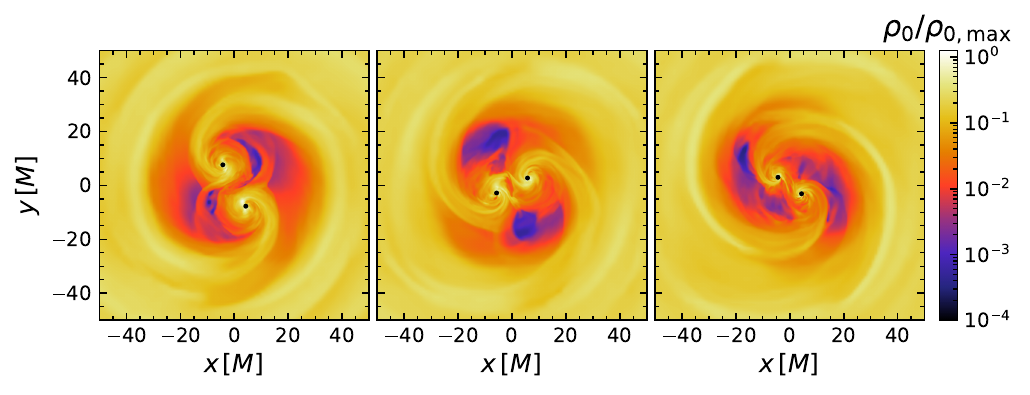}
        \caption{Contours of rest-mass density ($\rho_0$) normalized to the initial maximum density $\rho_{0, \rm max} =  2.6 \times 10^{-11} (\langle \dot{M}\rangle/0.1 \dot{M}_{\rm edd}) (M/10^7 M_{\odot})^{-1} (\eta/0.1)^{-1} \rm \, g \, cm^{-3}$, where $\dot{M}$ is the accretion rate and $\dot{M}_{\rm edd}$ is the Eddington accretion rate for a gravitational mass $M$ and radiative efficiency $\eta$. The left panel corresponds to the binary just after apocenter ($t/M=4120$), the center as it approaches pericenter ($t/M=4220$), and the right at pericenter ($t/M=4300$). We indicate the BH horizons with black disks. In the left panel, tidal streams circularize to form minidisks. In the center panel, these minidisks begin to accrete, and in the right panel are depleted at pericenter. Once in a quasi-steady-state, the binary exists in cavity with density $\rho_0/\rho_{0, \rm max} \sim 10^{-3}-10^{-2}$.}
        \label{fig:rho_panel}
    \end{figure*}

We initialize a gaseous torus around the BBH (as detailed in Section \ref{sec:approach}) and evolve the system until the accretion rate approximately relaxes (after $\sim 6$ orbits). In Figure~\ref{fig:rho_panel}, we plot the rest-mass density of the gas on the BBH orbital plane at representative times: at pericenter (left), at apocenter (right), and an intermediate time (middle). We indicate the BH apparent horizons with black disks. The gas density is normalized to the initial maximum gas density in the torus, $\rho_{0,\rm max}.$ The BHs continue to reside in a lower-density cavity ($\rho_0/\rho_{0, \rm max} \sim 10^{-3}$) within the higher-density circumbinary disk (CBD) ($\rho_0/\rho_{0, \rm max} \sim 1$). As the binary approaches the CBD inner edge, at its apocenter, it tidally torques the CBD and matter from the inner disk edge falls onto each BH through high-density tidal streams. The infalling gas temporarily circularizes around each BH creating a minidisk (left panel in Fig.~\ref{fig:rho_panel}), which begins to be depleted as the binary approaches the next pericenter passage (middle and right panels in Fig.~\ref{fig:rho_panel}). This process repeats quasi-periodically as the BBH inspirals.

    \begin{figure*}
        \includegraphics[width=\textwidth]{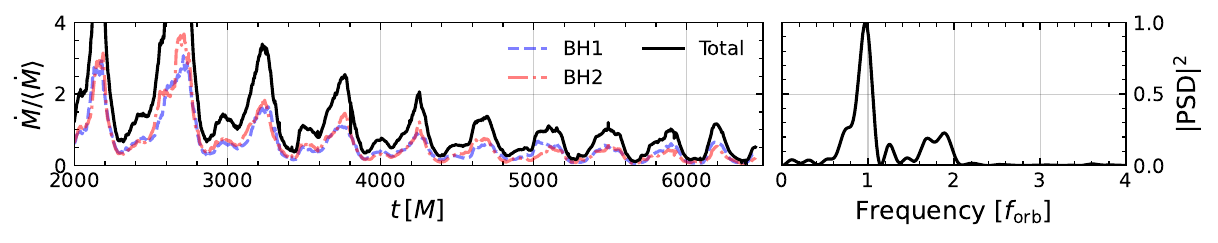}
        \caption{Left: Rest-mass accretion rates onto both BHs (solid black line), and onto the individual black holes (dashed lines), all normalized by the total average for $3000 < t/M < 5500$. Right: Power spectral density (PSD) of the Fourier transform of the total rest-mass accretion rate, with the frequency normalized to the BBH orbital frequency. The Fourier transform is performed on the time period $t=3000-5500 \, M$. The dominant frequency for accretion is $f\sim f_{\rm orb}$, and is insensitive to time interval over which we perform the Fourier transform.}
        \label{fig:mdot}
    \end{figure*}

In the left panel of Figure~\ref{fig:mdot}, we plot the time-dependent total rest-mass accretion rate onto the BHs (black solid line) and the accretion rate onto the individual black holes (dashed lines). Both exhibit quasiperiodic behavior. The mass accretion rate is initially high ($t/M<4000$) and decreases with time for two reasons: 1. there is a transient phase due to the initial data relaxation, and 2. the binary is inspiraling rapidly due to its higher initial eccentricity. Due to the combination of these two effects, the inner disk relaxes (as measured by the accretion rate) for $t/M>4000$. In particular, after $4000\,M$ the accretion rate reaches a quasi-steady-state, as indicated by its near constant amplitude and variability vs time.

In the right panel of Figure~\ref{fig:mdot}, we plot the Fourier transform of the total rest-mass accretion rate for the time range $3000 < t/M < 5500$, which corresponds to about 5 binary orbits. The Fourier plot demonstrates that the dominant frequency of accretion rate variability is the orbital frequency, $f \sim f_{\rm orb}$. This periodicity is consistent with recent Newtonian hydrodynamic studies of eccentric binaries~\citep{westernacher-schneider_multi-band_2022, delaurentiis_relativistic_2024} which are valid for substantially larger orbital separations. The measured accretion rate periodicity of $f_{\rm orb}$ in the eccentric case is fundamentally different from the periodicity in quasi-circular binaries, where the accretion rate is modulated at $1.4 f_{\rm orb}$ \citep{paschalidis_minidisk_2021, bright_minidisk}. The reported periodicity from the FFT is robust and remains at $1 f_{\rm orb}$ even if we choose to perform it for $t/M>4000$. The additional power we see in the PSD between $1-2 f_{\rm orb}$ is likely due to the initial relaxation of the fluid because it disappears if the Fourier transform is performed for $t>5000M$. We defer a detailed discussion of the dynamics and periodicity mechanisms in eccentric BBHs to a follow-up work~\citep{Manikantan_BBH_eccentric}.

\subsection{Jet Launching} \label{sec:jet}

     \begin{figure}
        \includegraphics[width=0.46\textwidth]{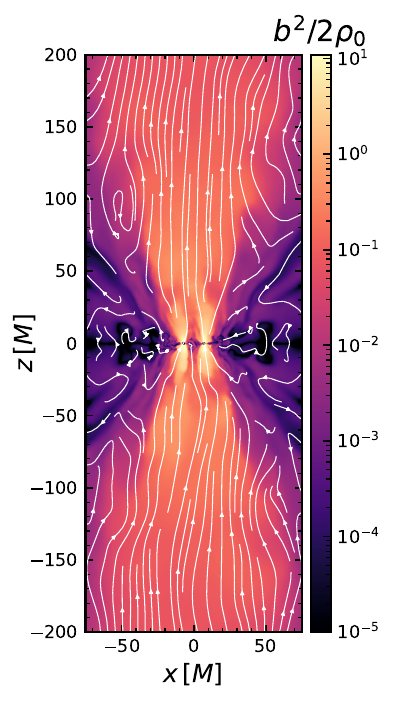}
        \caption{Contours of plasma magnetization ($\sigma\equiv b^2/2\rho_0$) on the $x-z$ plane at $t/M = 4197.6$. The BHs are at $x/M\pm 10$, $z/M=0$. Immediately below and above the BHs are regions of high plasma magnetization $\sigma \sim 10^{0}-10^{1}$ which extend vertically to $|z|> 200M$. To the left and right of the BHs is the CBD which has lower plasma magnetization $\sigma \sim 10^{-3}-10^{-5}$. The magnetic field (overplotted with directed white lines) is highly ordered above and below the BHs, indicating the jet regions.}\label{fig:mag}
    \end{figure}

We observe dual jet launching along the BBH orbital angular momentum axis consistent with previous CBD studies in full GR with non-spinning black holes~\citep{farris_binary_2012,Gold_PhysRevD.89.064060,gold_accretion_2014,khan_disks_2018,paschalidis_minidisk_2021, bright_minidisk}. In Figure~\ref{fig:mag}, we plot the plasma magnetization ($\sigma\equiv b^2/2\rho_0$) of our BBH-CBD system on the $x-z$ plane to describe the vertical dual-jet structure. The BHs exist on the $x-y$ plane ($z/M=0$) and launch jets in the $\hat{z}$-direction, which merge for $|z|/M\gtrsim 20$. These regions are magnetically dominated $\sigma > 1$ and extend out to $|z| > 200M$. Either side of the BBH, at $z/M \sim 0$, exists the CBD with low plasma magnetization $\sigma \sim 10^{-4}$. We show the magnetic field structure with directed white lines, which we plot over the plasma magnetization. In the jet regions above and below the BBH, the magnetic field is highly ordered and extends vertically out to $|z| > 200M$.

\section{Synchrotron Radiative Transfer} \label{sec:synchrotron}

    \begin{figure*}
        \includegraphics[width=\textwidth]{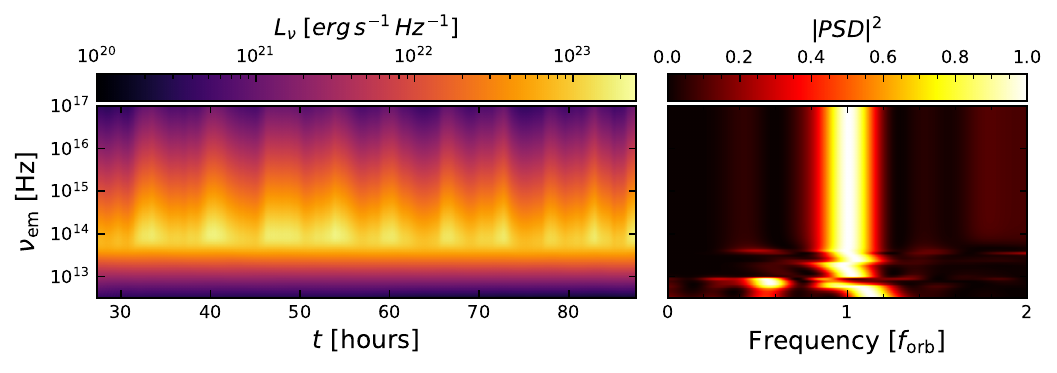}
        \caption{Left: specific luminosity of synchrotron emission on a color scale vs time for a $10^7 \, M_\odot$ binary accreting at $10\%$ Eddington. The $y$-axis is the frequency of the SED and the $x$-axis the time, with the color bar indicating the specific luminosity, $L_\nu$. Right: PSD of the Fourier transform of the frequency-binned SED time series. We perform the Fourier transform for the time period $t > 4000 \, M$ ($t > 54.7 \, \rm hours$). The plots demonstrate that only the optically thin region $\nu_{\rm em} \gtrsim 10^{14} \, \rm Hz$ of the synchrotron SED has clear periodicity on the orbital timescale, of $f \sim f_{\rm orb}$, which matches its mass accretion rate periodicity and Poynting luminosity periodicity. There is no clear periodicity in the optically thick regime. \label{fig:inu_fft}}
    \end{figure*}

To determine possible EM emission from these jet regions, we perform a synchrotron radiative transfer calculation of our simulations in postprocessing. In this section, we outline the key assumptions of our calculations and present spectral energy distributions (SEDs) of the jet synchrotron emission. We include a more detailed presentation of our synchrotron modeling and a derivation of the analytic scaling of the SED with the accretion rate and the BBH mass in a follow-up paper~\citep{Manikantan_BBH_eccentric}.

\subsection{Motivation and Set-up}

Relativistic electrons within the jet's magnetic field can produce synchrotron emission across the EM spectrum \citep[see, e.g.,][]{cheung2007, walker2018, kim2018, Saikia_2022JApA}. The interaction of the dual jets can give rise to current sheets and kink instabilities, which can then lead to a population of non-thermal electrons with power-law distribution~\citep{alves_efficient_2018, petropoulou_relativistic_2019,medina-torrejon_particle_2021,gutierrez_non-thermal_2024}. Motivated by this, we make the following approximations in our synchrotron modeling: 

\begin{enumerate}[noitemsep]
    \item We adopt a power-law electron distribution, assuming the local electron energy density equals 10\% of the local magnetic energy density in the jet~\citep{scott_low-frequency_1977, chevalier_synchrotron_1998, panaitescu_properties_2002,petropoulou_relativistic_2019}. This fraction is motivated by Fig.~6 in \cite{petropoulou_relativistic_2019} where the electron energy is $\sim10\%$ of the magnetic energy density; in reality, there is no widely accepted fraction to use here, but our reported periodicites of the synchrotron spectrum are independent of this fraction. We also tested a power-law distribution with minimum electron energy corresponding to a Lorentz factor of 2, as well as electrons with a thermal distribution~\citep{rybicki_lightman, tsouros_energy_2017}.
    \item We start our integrations at a height of $z/M = 50$ above the orbital plane. This is because we do not perform a general relativistic radiative transfer calculation, therefore our integration of the radiative transfer equation must be in approximately flat spacetime. However, we confirmed that the shape of the synchrotron spectrum and its variability are robust for integrations starting at $z/M=20$ and $z/M=30$.
    \item We adopt the "fast light" approximation; we solve the radiative transfer equation on a slice of constant coordinate time. This approximation is valid when the medium does not change much on a light crossing time, which in our case is valid for the optically thin synchrotron frequencies.
    \item We do not treat special relativistic effects other than those going into the computation of the synchrotron emissivity and absorption coefficients. This assumption is consistent with the fact that the plasma in the jet base in our simulations is only mildly relativistic ($v/c\sim 0.3$). 
\end{enumerate} 

In this work we choose the power in the electron power-law distribution to be $p=2.5$; in our follow-up extended paper, we will demonstrate that our reported results on the shape and the variability of the synchrotron spectrum are insensitive to $p$ and we will report the synchrotron spectra for quasi-circular and other eccentric binaries~\citep{Manikantan_BBH_eccentric}. Lastly, we only report results from a viewing angle of $\theta = 0$, in other words, we treat the system as a blazar. We solved the radiative transfer equation for other viewing angles and found that the shape and variability of the synchrotron spectrum are insensitive to the viewing angle for a given integration starting height. For the time-dependence of the synchrotron spectrum, we solve the radiative transfer equation with a cadence of $\sim 3.6 \, GM/c^3$ and produce SEDs for each slice of constant coordinate time. 

\subsection{Synchrotron Emission}

\begin{figure}
    \centering
    \includegraphics[scale=1]{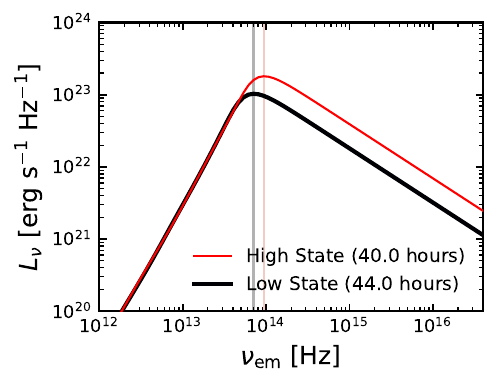}
    \caption{SED of the jet synchrotron emission at `low' (thick black line) and `high' (thin red line) states at $t = [44, 40]$ hours in Figure~\ref{fig:inu_fft}. The synchrotron self-absorption frequencies (vertical, translucent black and red lines) are $\nu_{\rm ssa} = [7.1, 9.3] \times 10^{13} \rm \, Hz$ and the peak specific luminosities are $L_{\nu} = [1.0, 1.8] \times 10^{23} \, \rm erg \, s^{-1} \, Hz^{-1}$ for the low and high states, respectively.}
    \label{fig:lnu_spectra}
\end{figure}

In the left panel of Figure~\ref{fig:inu_fft}, we plot the SED vs time, assuming a $10^7 \, M_\odot$ BBH accreting at $10\%$ Eddington, which is appropriate for our simulation of a thick accretion disk without full radiation transport. Given that we find very weak angular dependence of the specific intensity, to compute the specific luminosity we carry out the radiation transport from heights $z/M=50$ to $z/M = 200$ and assume the jet cross section (bounded by $\sigma \geq 0.1$) is emitting isotropically. The time-dependent SED is shown with a 2D color map where the $y$-axis indicates emission frequency and $x$-axis indicates increasing time in days, with the color indicating the specific luminosity. The frequency at peak synchrotron SED is $\nu_{\rm ssa} \sim 8\times 10^{13} \, \rm Hz$, where $\nu_{\rm ssa}$ is the synchrotron self-absorption frequency and oscillates between $7.1 < \nu_{\rm ssa}/10^{13} {\rm Hz} < 9.3$. This $\nu_{\rm ssa}$ is also a `break' frequency below which, in the optically thick regime, there is no clear time variability. Above $\nu_{\rm ssa}$, in the optically thin regime, the SED shows clear time variability in the form of `fringes', indicating periodically increased specific luminosity. The eccentric binary spends more time in a "low" state, where the synchrotron emission specific luminosity is minimum, than in a "high state", where the specific luminosity has a sharp rise and decay in time. In Figure~\ref{fig:lnu_spectra}, we plot the synchrotron SED at the low (thick black line) and high (thin red line) states, selected at times $t = [44, 40]$ hours from the left panel of Figure~\ref{fig:inu_fft}. We indicate the synchrotron self-absorption frequency $\nu_{\rm ssa}$ with translucent vertical lines for each of the SEDs, and note that the peak frequency shifts by $\sim 30\%$ between the low and high state occuring at $\nu_{\rm ssa} = [7.1, 9.3] \times 10^{13} \rm \, Hz$. Furthermore, the specific luminosity decreases from $L_{\nu} = 1.8 $ to $ 1.0\times 10^{23} \, \rm erg \, s^{-1} \, Hz^{-1}$ as the emission transitions from high to low. Our calculations predict that a smoking-gun synchrotron signature of more eccentric binaries is that they spend longer time in the low state than in the high state, and that the peak synchrotron frequency undergoes a shift between high and low states. This is consistent with eccentric binaries spending more time at apocenter than pericenter. 
     
In the right panel of Figure \ref{fig:inu_fft}, we show the Fourier transforms of the frequency-binned synchrotron SED. The $y$-axis is the frequency of the EM spectrum, and the $x$-axis is the frequency of the time variability (normalized by $f_{\rm orb}$) of the specific luminosity. The color map shows the strength of the power spectral density (PSD). The plot demonstrates that the synchrotron SED exhibits a periodicity on the orbital time $f \sim f_{\rm orb}$ for all EM frequencies in the optically thin regime (where $\nu_{\rm em} \gtrsim \nu_{\rm ssa}$) which is consistent with the dominant periodicity of its rest-mass accretion rate (Figure \ref{fig:mdot}). This is the first explicit demonstration that an EM signature periodicity matches the accretion rate periodicity in BBH accretion.

We also experimented with different ways to set the energy density in the electron power-law distribution. If we do not assume a fixed ratio between the electron and magnetic energy density, then variability is not as clear. Moreover, when we adopt a thermal distribution for the electrons the variability in the synchrotron SED also becomes inconclusive, i.e., there is no clear dominant frequency in the periodogram. However, we emphasize that a full general-relativistic ray-tracing and radiative transfer is necessary to determine the robustness of this finding.

\subsection{Coincident GW and EM Emission}\label{sec:coincident}

    \begin{figure}
        \centering
        \includegraphics[scale=0.925]{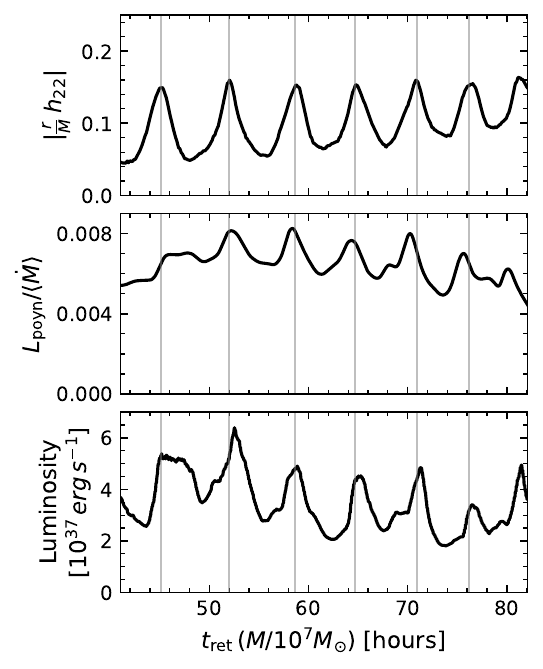}
        \caption{Top row: Amplitude of the $\ell=2, m=2$ mode of the gravitational wave (GW) strain, $\vert h_{22}\vert $, normalized by the distance $r/M$ vs retarded time $t_{\rm ret}$. Middle: Outgoing Poynting luminosity normalized by the rest-mass accretion rate vs $t_{\rm ret}$. Bottom: synchrotron luminosity integrated in the optically thin regime $8\times 10^{13} Hz < \nu_{\rm em} < 8\times 10^{14} Hz$ vs $t_{\rm ret}$. We denote the location of GW bursts with vertical translucent gray lines on both the GW and EM panels. The lightcurves shows almost perfect alignment between GW, Poynting flux, and EM bursts.}
        \label{fig:gw_em}
    \end{figure}

In Figure \ref{fig:gw_em}, we showcase the coincident GW and EM emission from our SMBBH. We plot the amplitude of the $\ell=2, m=2$ mode of the GW strain (top panel), the outgoing Poynting luminosity normalized by the time-averaged rest-mass accretion rate (middle panel), and the optically thin synchrotron luminosity (bottom panel) integrated for frequencies $\nu_{\rm em} \in [8\times 10^{13}, 8\times 10^{14}] \rm \, Hz$ (which sample the SED near the peak) all vs retarded time. We indicate the GW peaks with vertical translucent lines in all three panels. The figure demonstrates that the GW and EM bursts happen almost simultaneously -- the GW bursts marginally precede the EM bursts in certain cases (see the second, third, and fourth peaks). A smoking-gun multi-messenger signature of binaries with non-negligible eccentricity is that the time period between successive bursts is the same for both the GW and EM synchrotron emission.

\subsection{Detectability of Synchrotron Emission}

The synchrotron emission we report can be detected by NIRCam and MIRI on the James Webb Space Telescope~\citep{greene_2017, miri-intro, miri-inflight}, the upcoming Rubin Observatory and the Legacy Survey of Space and Time (LSST), as well as the Roman Space Telescope~\citep{ivezic_lsst_2019}. The sensitivity of these instruments places limits on the distance of the detectability of the synchrotron signatures we predict. We can solve for the maximum observable distance of our objects by starting with the equation for the AB magnitude of an object~\citep{oke_secondary_1983, bessell_standard_2005},
\begin{equation}
    m = -2.5 \log_{10}{\frac{F_\nu}{F_0}},
\end{equation}
where $m$ is the AB magnitude, $F_\nu$ is the specific flux of the object, and $F_0=3631 \rm \, Jy$ is the reference flux, where $1 \, {\rm Jy} = 10^{-23} \, \rm erg \, s^{-1} \, cm^{-2} \, Hz^{-1} $. The flux of our object can be expressed in terms of its specific luminosity,
\begin{equation}
    F_\nu = L_\nu / 4\pi d^2,
\end{equation}
where $L_\nu$ is the specific luminosity of our synchrotron radiation and $d$ is the luminosity distance to the object. Solving for the distance we obtain
%

%
\begin{equation}
    d = \sqrt{\frac{L_\nu}{4 \pi F_0}10^{m/2.5}}.
\end{equation}

The most sensitive filter on NIRCam (F150W2) has a limiting magnitude of $m=29.8$ for a $10^4 \rm \, second$ exposure with a signal-to-noise ratio of $10$~\citep{greene_2017}. Putting this into the equation, along with our calculated optically thin synchrotron specific luminosities, we can estimate the maximum distance our BBH could be observed out to. For example, we estimate that NIRCam could observe our $10^7 M_{\odot}$ BBH with peak $L_{\nu} \sim 10^{23} \, \rm erg \, s^{-1} \, Hz^{-1}$ (see Figure \ref{fig:inu_fft}) out to $\sim 0.2 \, \rm Gpc$, which corresponds to a redshift $z\sim 0.04$ assuming standard $\Lambda$CDM cosmology. We also repeated the calculation for a $10^9 M_{\odot}$ BBH with peak $L_\nu \sim 10^{26} \, \rm erg \, s^{-1} \, Hz^{-1}$, which would be detectable out to $\sim 7 \, \rm Gpc$ ($z\sim 1.1$), although, at these redshifts and binary mass, the optically thin synchrotron frequency would be redshifted into the F322W2 NIRCam filter, which has a lower limiting magnitude of $m=29.1$, and MIRI filters, which, for its most sensitive filter (F560W) has a limiting magnitude of $m=26.2$~\citep{wright_cosmology_2006, greene_2017}. Therefore, when evaluating more massive candidates at higher redshift, a more careful calculation of the detectability is needed. All these distances are $\sim 5\times$ greater if we integrate for $z\geq 20 \, M$ and the peak frequencies are blueshifted. Therefore, the above estimates are preliminary and a more detailed radiative transfer calculation with ray tracing is necessary for more precise distances.

Additionally, we can roughly estimate the thermal emission of our CBD and minidisks. The inner parts of such accretion disks are radiation pressure dominated and optically thick due to Thompson scattering~\citep{shapiro_black_1983}. Thompson scattering increases the path length of photons before they escape which increases the photon's probability of absorption and reemission. This results in thermalization of the EM emission - we refer the reader to \cite{shapiro_black_1983} for a more detailed discussion. Therefore, using the fluid pressure, $P$, from our simulations, modeling it as $P=\frac{1}{3}aT^4$ (consistent with our choice of $\Gamma=4/3$), and assuming that accretion takes place at $10\%$ of the Eddington rate, we can solve for the temperature of the CBD and minidisk to estimate the ion temperature. We find this temperature to be $\mathcal{O}({10^{6}}\, \rm K)$. With corrections to the effective temperature and specific flux (Eqs.~(14.5.54) and (14.5.51) from \cite{shapiro_black_1983}, respectively) and integrating over the photosphere, we estimate the peak thermal emission frequency to be $\sim 10^{15} \rm \, Hz$ with a peak specific flux of $F_{\rm bb} \sim 10^{24} \rm \, erg \, s^{-1} \, Hz^{-1}$. This is more luminous than our peak synchrotron emission at $\nu_{\rm ssa}$ if our radiative transfer integration starts at $z/M=50$. However, if we start our synchrotron radiation transfer at $z/M=20$, the `high' states of our synchrotron timeseries become more luminous than the predicted black body emission from the entire disk and minidisks. Therefore, we tentatively predict the jet synchrotron emission to be observable over the disk thermal spectrum. We reiterate, however, that a more careful calculation of the jet, CBD, and minidisk emission with a full general-relativistic ray tracing and radiative transfer is necessary to more reliably decipher emission from these systems.

\section{Conclusions} \label{sec:conc}

In this work, we presented the first simulation of magnetohydrodynamic disk accretion onto an equal-mass, non-spinning, eccentric binary black hole in $3+1$ full general relativity, incorporating synchrotron emission through the jet in post processing for the first time. Our key findings are: 
\begin{enumerate}[noitemsep]
    \item The accretion rate onto eccentric BBHs in the strong-field dynamical spacetime regime has periodicity that matches the binary orbital frequency $f \sim f_{\rm orb}$, unlike quasi-circular binaries for which this occurs at $f\sim 1.4 \, f_{\rm orb}$.
    \item Eccentric binaries exhibit periodicity in the Poynting luminosity and optically thin synchrotron emission at the orbital frequency, $f_{\rm orb}$. A smoking-gun signature of eccentric binaries is that they spend more time in a low state (lower luminosity) than in a high state (higher luminosity) consistent with the time spent at apocenter and pericenter. Moreover, the peak Synchrotron frequency shifts between the high and low states by $\sim 30\%$.
    \item A smoking gun multimessenger signature of eccentric binaries is quasiperiodic bursts in their GWs, optically thin synchrotron emission, and Poynting luminosity, with identical delay times for consecutive EM and GW bursts.
\end{enumerate}

We also find that the synchrotron emission variability is sensitive to the choice of electron energy distribution - only a powerlaw electron distribution with a fixed ratio between electron and magnetic energy density demonstrates clear variability. Assuming no equipartition or a thermal electron distribution does not reveal a periodic synchrotron source. This could offer a unique opportunity to probe jet plasma physics with multimessenger observations with GWs: LISA BBHs that are found to have substantial eccentricity ($O(0.1$)) can be followed up with EM observations to test jet variability. The existence or not of variability of the synchrotron emission from the jet base in conjunction with theoretical modeling such as that performed in this work, can allow us to understand how electrons adapt to the changing magnetic field in a BBH spacetime. For example,  lack of variability at the binary orbital period would inform us that either the electrons near the jet base do not follow the magnetic field energy density or that they follow a thermal distribution. Regardless, the variability in the Poynting luminosity is robust, which implies that as the jet propagates into the interstellar and intergalactic medium it could generate emission in the radio that will exhibit variability on the binary orbital time. 

We conclude by listing some caveats: a fully general-relativistic ray-tracing and radiative transfer calculation is necessary to decipher the exact emission from the jet region down to the BH horizons and to understand EM emission from the circumbinary disk, the inner cavity, and the minidisks. The latter is expected to be responsible for the bulk of the X-ray/UV emission as well as Doppler shifted emission lines \citep{sesana_multimessenger_2012, charisi_multi-messenger_2022, bogdanovic_bhb_review}. Radiation feedback becomes important for accretion rates near the Eddington regime and must be accounted for. As such, our results are most applicable to sub-Eddington SMBBHs. Finally, it is important to consider a wide range of disk initial data and include spinning BHs as well as other values of mass ratio and eccentricity. These will be the subject of future works of ours. However, the periodicities driven by eccentricity that we report should be robust. Hence, the qualitative features we have discovered in this work should be invariant under the aforementioned caveats as long as the eccentricity is high enough.

\begin{acknowledgments}
    We thank Collin Christy for many useful discussions. This work was in part supported by NASA grant 80NSSC24K0771 and NSF grant PHY-2145421 to the University of Arizona. This research is part of the Frontera computing project at the Texas Advanced Computing Center. Frontera is made possible by U.S. National Science Foundation award OAC-1818253. This work used Stampede2 and Stamepede3 at the Texas Advanced Computing Center through allocation PHY190020 from the Advanced Cyberinfrastructure Coordination Ecosystem: Services \& Support (ACCESS) program, which is supported by U.S. National Science Foundation grants 2138259, 2138286, 2138307, 2137603, and 2138296~\citep{boerner_access_2023}.
\end{acknowledgments}

\bibliography{sample631}{}
\bibliographystyle{aasjournal}

\end{document}